\documentclass[twocolumn,aps,superscriptaddress,showpacs]{revtex4}
\usepackage{amssymb}
\usepackage{amsmath,bm}
\usepackage{graphicx}
\usepackage[normalem]{ulem}
\usepackage[dvips]{color}

\renewcommand\sout{\bgroup \color{red} \ULdepth=-.5ex \ULset}

\begin{document}

\title{Constraining the density slope of nuclear symmetry energy at subsaturation
densities using electric dipole polarizability in $^{208}$Pb}
\author{Zhen Zhang}
\affiliation{Department of Physics and Astronomy and Shanghai Key Laboratory for
Particle Physics and Cosmology, Shanghai Jiao Tong University, Shanghai 200240, China}
\author{Lie-Wen Chen\footnote{%
Corresponding author (email: lwchen$@$sjtu.edu.cn)}}
\affiliation{Department of Physics and Astronomy and Shanghai Key Laboratory for
Particle Physics and Cosmology, Shanghai Jiao Tong University, Shanghai 200240, China}
\affiliation{Center of Theoretical Nuclear Physics, National Laboratory of Heavy Ion
Accelerator, Lanzhou 730000, China}
\date{\today}

\begin{abstract}
Nuclear structure observables usually most effectively probe the properties of
nuclear matter at subsaturation densities rather than at saturation density.
We demonstrate that the electric dipole polarizibility $\alpha _ {\text{D}}$ in
$^{208}$Pb is sensitive to both the magnitude $E_{\text{sym}}(\rho_{\text{c}})$
and density slope $L(\rho_{\text{c}})$ of the symmetry energy at the subsaturation
cross density $\rho_{\text{c}} = 0.11$ fm$^{-3}$. Using
the experimental data of $\alpha _ {\text{D}}$ in $^{208}$Pb from RCNP and the
recent accurate constraint of $E_{\text{sym}}(\rho_{\text{c}})$ from
the binding energy difference of heavy isotope pairs, we extract a value of
$L(\rho_{\text{c}}) = 47.3 \pm 7.8$ MeV. The implication of the present
constraint of $L(\rho_{\text{c}})$ to the symmetry energy at saturation density,
the neutron skin thickness of $^{208}$Pb and the core-crust transition density
in neutron stars is discussed.
\end{abstract}

\maketitle

\section{Introduction}

Due to its multifaceted roles in nuclear physics and
astrophysics~\cite{Lat04,Ste05,Bar05,LCK08} as well as new physics
beyond the standard model~\cite{Hor01b,Sil05,Wen09,Zhe14}, the
symmetry energy has become a hot topic in current research frontiers
of nuclear physics and astrophysics~\cite{LiBAEPJA14}. During the
last decade, a lot of experimental, observational and theoretical
efforts have been devoted to constraining the magnitude $E_{\text{sym}}(\rho)$
and density slope $L(\rho)$ of the symmetry energy at nuclear
saturation density $\rho_0$ ($\sim 0.16$ fm$^{-3}$), i.e.,
$E_{\text{sym}}(\rho_0)$ and $L(\rho_0)$. Although important
progress has been made, large uncertainties on the values of
$E_{\text{sym}}(\rho_0)$ and $L(\rho_0)$ still exist (See, e.g.,
Refs.~\cite{Bar05,Ste05,LCK08,LiBAEPJA14,Tsa12,Lat12,ChenLW12,LiBA12,Hor14}).
For instance, while the $E_{\text{sym}}(\rho_0)$ is determined to be
around $32\pm 4$ MeV, the extracted $L(\rho_0)$ varies significantly
from about $20$ to $115$ MeV, depending on the observables and analysis
methods. To better understand the model dependence and narrow the
uncertainties of the constraints is thus of extreme importance.

While many studies on heavy ion collisions and neutron stars have
significantly improved our knowledge on the symmetry energy, more
and more constraints on the symmetry energy have been obtained in
recent years from analyzing the properties of finite nuclei, such
as the nuclear binding energy~\cite{Mye96,Dan09,Liu10,Mol12,LatX12},
the neutron skin thickness~\cite{War09,Che10,Vin14}, and the resonances and
excitations~\cite{Sag07,Kli07,Car10,Rei10,Tam11,Pie12,Roc13,Pie14,Col14}.
Furthermore, it has been realized that the properties of finite nuclei
usually provide more precise constraints on $E_{\text{sym}}(\rho)$ and
$L(\rho)$ at subsaturation densities rather than at saturation density
$\rho_0$. This feature is understandable since the characteristic
(average) density of finite nuclei is less than $\rho_0$. For example,
the average density of heavy nuclei (e.g., $^{208}$Pb) is about
$0.11$ fm$^{-3}$, and thus the properties of heavy nuclei most
effectively probe the properties of nuclear matter around
$0.11$ fm$^{-3}$~\cite{Tri08,Cao08,Cen09,Che11,Kha12,Zha13,Bro13,Wan13,Roc13a,Dan14,Fat14}.
Indeed, a quite accurate constraint on the symmetry energy at the
subsaturation cross density $\rho_{\text{c}} = 0.11$ fm$^{-3}$, i.e.,
$E_{\text{sym}}(\rho_{\text{c}})=26.65 \pm 0.20$MeV, has been recently
obtained from analyzing the binding energy difference of heavy isotope
pairs~\cite{Zha13}. In contrast to the fact that many and precise
constraints on the magnitude of $E_{\text{sym}}(\rho )$ around $\rho_{\text{c}}$ have
been obtained, to the best of our knowledge, so far there is only one
experimental constraint on the density slope $L(\rho_{\text{c}})$ which was obtained from
analyzing the neutron skin data of Sn isotopes~\cite{Zha13}. Knowledge on
$L(\rho_{\text{c}})$ is not only important for understanding
the density dependence of the symmetry energy itself, but also plays a
central role in determining the neutron skin thickness of heavy nuclei
and the core-crust transition density in neutron stars. Therefore, any
new constraints on $L(\rho_{\text{c}})$ will be extremely useful.

In the present work, with the precise knowledge of
$E_{\text{sym}}(\rho_{\text{c}})$, we demonstrate that the
electric dipole polarizability $\alpha_{\text{D}}$ in $^{208}$Pb measured
at the Research Center for Nuclear Physics (RCNP) via polarized proton
inelastic scattering at forward angles, can put a strong limit on the
$L(\rho_{\text{c}})$. We emphasize since at forward angles Coulomb
excitation dominates, the extracted $\alpha_{\text{D}}$ at RCNP is
expected to be a relatively clean isovector indicator with less
uncertainties from strong interaction.

\section{Model and method}

\subsection{The symmetry energy and Skyrme-Hartree-Fock approach}

The equation of state (EOS) of asymmetric nuclear matter, given by its binding
energy per nucleon, can be written as
\begin{equation}
E(\rho ,\delta )=E_{0}(\rho )+E_{\mathrm{sym}}(\rho )\delta ^{2}+O(\delta
^{4}),  \label{EOSANM}
\end{equation}%
where $\rho $ is the baryon density, $\delta=(\rho _{n}-\rho _{p})/(\rho _{p}+\rho _{n})$
is isospin asymmetry, $E_{0}(\rho )=E(\rho ,\delta =0)$ is the EOS of
symmetric nuclear matter, and the symmetry energy is expressed as
\begin{equation}
E_{\mathrm{sym}}(\rho )=\frac{1}{2!}\frac{\partial ^{2}E(\rho ,\delta )}{%
\partial \delta ^{2}}|_{\delta =0}.  \label{Esym}
\end{equation}%
Around a reference density $\rho _{r}$, the $E_{\mathrm{sym}}(\rho )$ can
be expanded in $\chi_r=(\rho -{\rho _{r}})/\rho _{r}$ as
\begin{equation}
E_{\text{sym}}(\rho )=E_{\text{sym}}({\rho _{r}})+\frac{L(\rho _{r})}{3}\chi_r+O(\chi_r^2),
\end{equation}
where $L(\rho _{r})=3{\rho _{r}}\frac{\partial E_{\mathrm{sym}}(\rho )}{\partial
\rho }|_{\rho ={\rho _{r}}}$ is the density slope parameter which characterizes
the density dependence of the symmetry energy around $\rho _{r}$.

Our calculations in the present work are based on the Skyrme-Hartree-Fock (SHF)
approach with the so-called standard Skyrme force (see, e.g.,
Ref.~\cite{Cha97,Fri86,Klu09}) which includes $10$ parameters, i.e., the $9$
Skyrme force parameters $\sigma $, $t_{0}-t_{3}$, $x_{0}-x_{3}$, and the
spin-orbit coupling constant $W_{0}$. Instead of directly using the $9$ Skyrme
force parameters, we can express them explicitly in terms of $9$ macroscopic
quantities, i.e., $\rho _{0}$, $E_{0}(\rho_{0})$, the incompressibility $K_{0}$,
the isoscalar effective mass $m_{s,0}^{\ast }$, the isovector effective mass
$m_{v,0}^{\ast }$, $E_{\text{sym}}({\rho _{r}})$, $L({\rho _{r}})$, the gradient
coefficient $G_{S}$, and the symmetry-gradient coefficient $G_{V}$. In this case,
we can examine the correlation of properties of finite nuclei with each individual
macroscopic quantity by varying individually these macroscopic quantities within
their empirical ranges. Recently, this correlation analysis method has been
successfully applied to study nuclear matter properties from analyzing nuclear
structure observables~\cite{Che10,Che11a,Che11,Che12,Zha13}, and will also be
used in this work.

\subsection{Random-phase approximation and electric dipole polarizability}

The random-phase approximation (RPA) provides an important microscopic
approach to calculate the electric dipole polarizability in finite nuclei.
Within the framework of RPA theory, for a given excitation operator $\hat{F}_{JM}$,
the reduced transition probability from RPA ground state $|\tilde{0}\rangle $
to RPA excitation state $|\nu \rangle  $ is given by:
\begin{equation}
\begin{split}
B(EJ:\tilde{0}\rightarrow|\nu\rangle)&=|\langle\nu||\hat{F}_{J}||\tilde{0}\rangle|^2\\
&=\left|\sum_{mi}\left( X_{mi}^{\nu}+Y_{mi}^{\nu}\right) |\langle m||\hat{F}_{J}||i\rangle\right| ^2
\end{split},
\end{equation}
where $m (i)$ denotes the unoccupied (occupied) single nucleon state;
$\langle m||\hat{F}_{J}||i\rangle$ is the reduced matrix element of $\hat{F}_{JM}$;
and $X_{mi}^{\nu}$ and $Y_{mi}^{\nu}$ are the RPA amplitudes. The strength
function then can be calculated as:
\begin{equation}
S(E)=\sum_{\nu}|\langle\nu\Vert\hat{F}_J\Vert\tilde{0}\rangle|^2\delta(E-E_{\nu}),
\end{equation}
where $E_{\nu}$ is the energy of RPA excitation state $|\nu\rangle$. Thus the
moments of strength function can be obtained as:
\begin{equation}
m_k=\int dE E^kS(E)=\sum_{\nu}|\langle\nu\Vert\hat{F}_J\Vert\tilde{0}\rangle|^2E_{\nu}^k.
\end{equation}
In the case of electric dipole ($E1$) response, the excitation operator is defined as:
\begin{equation}
\hat{F}_{1M} = \frac{eN}{A}\sum^Z_{i=1}r_iY_{\text{1M}}(\hat{r}_i)-\frac{eZ}{A}\sum^N_{i=1}r_iY_{\text{1M}}(\hat{r}_i),
\end{equation}
where $Z$, $N$ and $A$ are proton, neutron and mass number, respectively;
$r_i$ is the nucleon's radial coordinate; $Y_{\text{1M}}(\hat{r_i})$ is the corresponding
spherical harmonic function. For a given Skyrme interaction, we can calculate
the inverse energy-weighted moment $m_{-1}$ using the HF-RPA method, and then obtain
the electric dipole polarizability $\alpha_{D}$ as
\begin{equation}
\alpha_{D}=\frac{8\pi}{9}e^2\int dEE^{-1}S(E)=\frac{8\pi}{9}e^2m_{-1}. \label{AlphaDMm1}
\end{equation}

\begin{figure*}[tbp]
\includegraphics[scale=0.5]{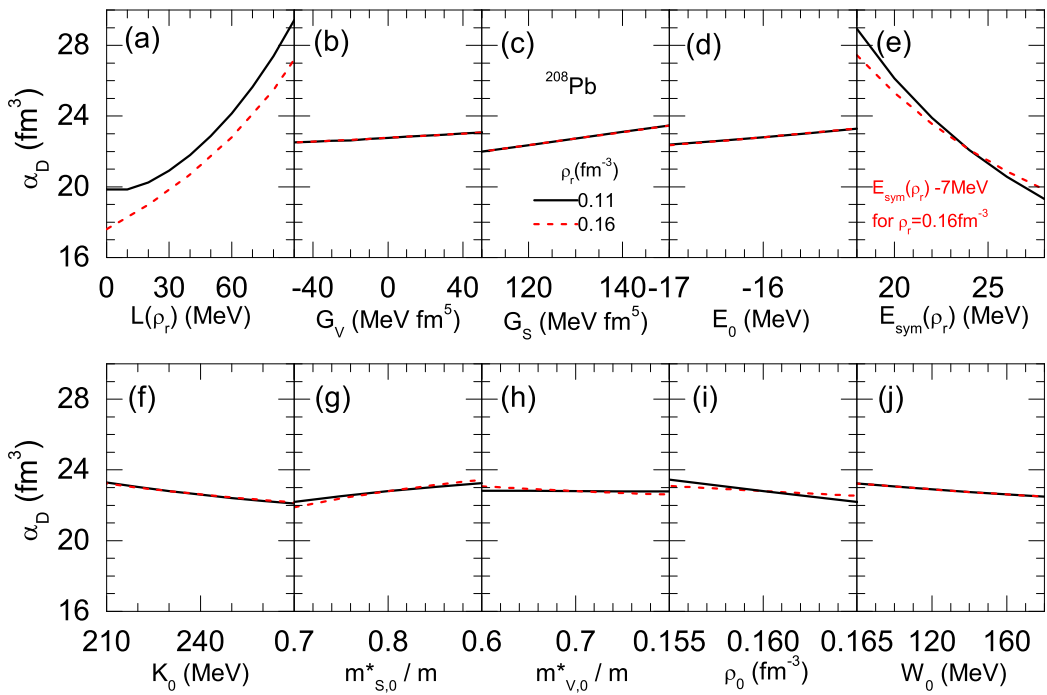}
\caption{(Color online) The electric dipole polarizability $\alpha_{\text{D}}$
in $^{208}$Pb from SHF-RPA calculations with the MSL0 interaction by varying
individually $L(\rho_{\text{r}})$ (a), $G_{V}$ (b), $G_{S}$ (c), $E_{0}(\rho _{0})$ (d),
$E_{\text{sym}}(\rho _{\text{r}})$ (e), $K_{0}$ (f), $m_{s,0}^{\ast }$ (g),
$m_{v,0}^{\ast }$ (h), $\rho _{0}$ (i), and $W_{0}$ (j) for
$\rho_{\text{r}}=0.11$ and $0.16{~\text{fm}}^{-3}$. The $E_{\mathrm{sym}}(\rho_r)$ is shifted by subtracting 7 MeV for $\rho_r=0.16{~\text{fm}}^{-3}$.}
\label{AlphaD}
\end{figure*}

For the theoretical calculations of electric dipole polarizability in
$^{208}$Pb in the present work, we employ the Skyrme-RPA program by
Col$\grave{\text{o}}$ {\it et al}~\cite{Colo13}. In this program, the
SHF mean field and the RPA excitations are fully self-consistent. In
particular, we calculate the isovector dipole strength in $^{208}$Pb
with a spherical box extending up to $24$ fm, a radial mesh of $0.1$ fm
and a cutoff energy of $E_{\text{C}} = 150$ MeV which denotes the
maximum energy of the unoccupied single-particle states in the RPA
model space. Then the inverse energy-weighted moment $m_{-1}$ is evaluated
with an upper integration limit of $130$ MeV according to the experimental
energy range~\cite{Tam11}, and thus the electric dipole polarizability
$\alpha_{\text{D}}$ can be calculated invoking Eq.~\eqref{AlphaDMm1}.

\subsection{The symmetry energy and electric dipole polarizability}
The electric dipole polarizability $\alpha_{\text{D}}$ has been shown
to be a sensitive probe of the symmetry energy~\cite{Rei10,Pie12,Roc13}.
In particular, based on the droplet model, Roca-Maza {\it et al.}~\cite{Roc13}
obtained the following relation:
\begin{equation}
\label{AlphaDDM}
\alpha_{\mathrm{D}}= \frac{\pi e^2}{54}
\frac{A\left\langle r^2 \right\rangle}{E_{\mathrm{sym}}(\rho_0)}
\left[1+\frac{5}{3}
\frac{E_{\mathrm{sym}}(\rho_0)-a_{\mathrm{sym}}(A)}{E_{\mathrm{sym}}(\rho_0)}
\right],
\end{equation}
where $\langle r^2 \rangle$ is the mean-square radius and $a_{\mathrm{sym}}(A)$
is the symmetry energy coefficient of a finite nucleus of mass number $A$. Furthermore,
using the empirical relation $a_{\mathrm{sym}}(A)\approx E_{\mathrm{sym}}(\rho_A)$~\cite{Cen09,Che11,Dan14}
and expanding $E_{\mathrm{sym}}(\rho_A)$ as
\begin{equation}
 E_{\mathrm{sym}}(\rho_A)\approx E_{\mathrm{sym}}(\rho_0)-L(\rho_0)(\rho_0-\rho_A)/3\rho_0,
\label{EsymrA}
\end{equation}
Roca Maza \textit{et al.} demonstrated that $\alpha_{\mathrm{D}}$ is correlated
with both $E_{\mathrm{sym}}(\rho_0)$ and $L(\rho_0)$. Particularly, based on a
large and representative set of relativistic and nonrelativistic nuclear
mean-field models, they found a strong linear correlation between
$\alpha_{\mathrm{D}}E_{\mathrm{sym}}(\rho_0)$ and $L(\rho_0)$ and then extracted the
constraint $L(\rho_0) = 43 \pm (6)_{\text{expt}}\pm(8)_{\text{theor}}\pm(12)_{\text{est}}$ MeV
from the combination of the experimental determination of $\alpha_{\mathrm{D}}$ with the
empirical estimate of $E_{\text{sym}}(\rho_0) = 31 \pm (2)_{\text{est}}$ MeV. One can see that
the uncertainty of the estimated $E_{\mathrm{sym}}(\rho_0)$ leads to a large error of
$12$ MeV for $L(\rho_0)$.

Instead of expressing $E_{\mathrm{sym}}(\rho_A)$ in terms of $E_{\mathrm{sym}}(\rho_0)$
and $L(\rho_0)$ as in Eq.~(\ref{EsymrA}), one can also express $E_{\mathrm{sym}}(\rho_0)$
in terms of $E_{\mathrm{sym}}(\rho_c)$ and $L(\rho_c)$ as
\begin{equation}
 E_{\mathrm{sym}}(\rho_0)\approx E_{\mathrm{sym}}(\rho_c)+L(\rho_c)(\rho_0-\rho_c)/3\rho_c.
\label{Esymr0rc}
\end{equation}
Noting $\rho_{208}\approx \rho_c$~\cite{Cen09,Che11,Dan14}, one can then see from
Eqs.~(\ref{AlphaDDM}) and (\ref{Esymr0rc}) that $\alpha_{\mathrm{D}}$ in $^{208}$Pb
is also correlated with both $L(\rho_c)$ and $E_{\mathrm{sym}}(\rho_c)$. As we
will see in the following, the microscopic RPA calculations indeed show that $\alpha_D$ is
sensitive to $E_{\mathrm{sym}}(\rho_0)$ and $L(\rho_0)$ as well as to $L(\rho_c)$
and $E_{\mathrm{sym}}(\rho_c)$. Since $E_{\mathrm{sym}}(\rho_c)$ has been
stringently constrained recently (see, e.g., $E_{\mathrm{sym}}(\rho_c) = 26.65\pm0.20$
MeV in Ref.\citep{Zha13}), the $\alpha_{\mathrm{D}}$ in $^{208}$Pb can thus be used to
constrain the $L(\rho_c)$ parameter.

\section{Results and discussions}

To examine the correlation of the $\alpha_{\text{D}}$ in $^{208}$Pb with each
macroscopic quantity, especially on $E_{\text{sym}}({\rho _r})$
and $L(\rho_r)$, we show in
Fig.~\ref{AlphaD} the $\alpha_{\text{D}}$ in $^{208}$Pb from SHF with the
Skyrme force MSL0~\cite{Che10} by varying individually $L(\rho_r)$,
$G_{V}$, $G_{S}$, $E_{0}(\rho _{0})$, $E_{\text{sym}}(\rho _r)$,
$K_{0}$, $m_{s,0}^{\ast }$, $m_{v,0}^{\ast }$, $\rho _{0}$, and $W_{0}$ within
their empirical uncertain ranges, namely, varying one quantity at a time
while keeping all others at their default values in MSL0, for $\rho_r=0.11$
and $0.16$ fm$^{-3}$, respectively. It is seen from Fig.~\ref{AlphaD} that,
as Eq.~(\ref{AlphaDDM}) suggests, the $\alpha_{D}$ in $^{208}$Pb exhibits strong
correlations with both $L(\rho_{\text{r}})$ and $E_{\text{sym}}(\rho _{\text{r}})$,
while much weaker correlation with other macroscopic quantities. Particularly, the
$\alpha_{\text{D}}$ decreases sensitively with $E_{\text{sym}}(\rho_r)$
while increases rapidly with $L({\rho_r})$, implying a fixed value of
$\alpha_{\text{D}}$ will lead to a strong positive correlation between
$E_{\text{sym}}(\rho_r)$ and $L({\rho_r})$. The results for $\rho_r = 0.16$ fm$^{-3}$
just confirm the correlations of $\alpha_{\mathrm{D}}$ with $E_{\mathrm{sym}}(\rho_0)$
and $L(\rho_0)$ reported in Ref.\cite{Roc13}. For $\rho_r=0.11$ fm$^{-3}$, given
that the symmetry energy at $\rho_{\text{c}}=0.11$ fm$^{-3}$ has been well constrained
as $E_{\text{sym}}(\rho_{\text{c}})=26.65\pm0.20$ MeV, one thus expects the
$\alpha_{\text{D}}$ in $^{208}$Pb can constrain stringently
the parameter $L({\rho_{\text{c}}})$.

\begin{figure}[tbp]
\includegraphics[scale=0.35]{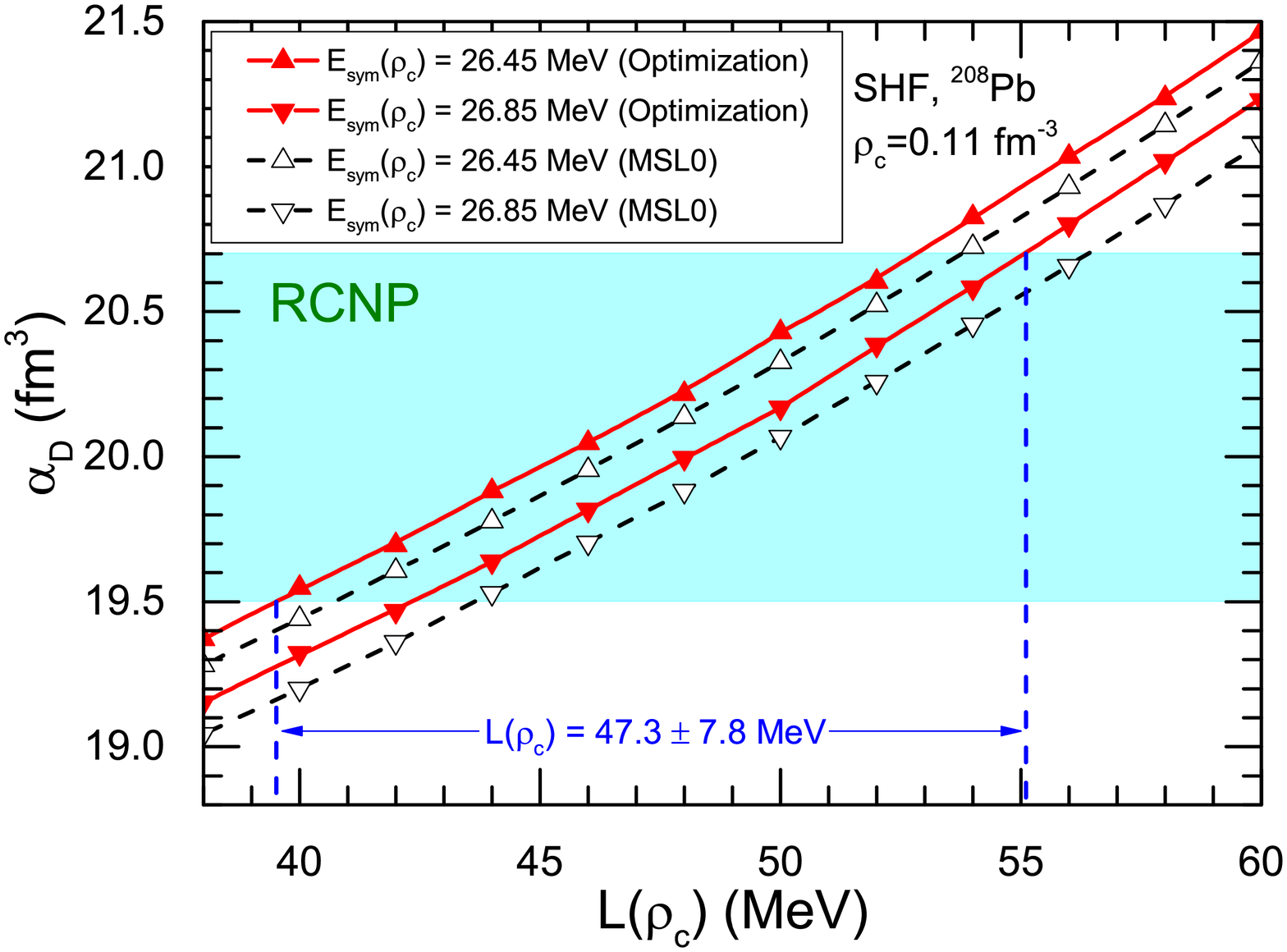}
\caption{(Color online) The electric dipole polarizability $\alpha_{\text{D}}$
in $^{208}$Pb as a function of $L(\rho_{\text{c}})$ for fixed
$E_{\text{sym}}(\rho_{\text{c}})$. The open (solid) up- and down-triangles
represent the results with $E_{\text{sym}}(\rho_{\text{c}}) = 26.45$ and
$26.85$ MeV, respectively, from SHF-RPA calculations with the values of other
parameters fixed in MSL0 (obtained in optimization). The band indicates the
experimental value of $\alpha_D = 20.1\pm 0.6$ fm$^3$ from RCNP~\cite{Tam11}.}
\label{AlphadLc}
\end{figure}

Fixing the values of other $8$ macroscopic quantities, i.e., $G_{V}$, $G_{S}$,
$E_{0}(\rho _{0})$, $K_{0}$, $m_{s,0}^{\ast }$, $m_{v,0}^{\ast }$, $\rho _{0}$
and $W_{0}$ at their default values in MSL0, we illustrate in Fig.~\ref{AlphadLc}
by open up-triangles (down-triangles) the $\alpha_{\text{D}}$ in $^{208}$Pb as a function of
$L(\rho_{\text{c}})$ for $E_{\text{sym}}(\rho_{\text{c}})=26.45~(26.85)$ MeV. As
expected, it is seen from Fig.~\ref{AlphadLc} that the $\alpha_{\text{D}}$ in $^{208}$Pb
increases (decreases) with $L({\rho_{\text{c}}})$ ($E_{\text{sym}}(\rho_{\text{c}})$)
for a fixed $E_{\text{sym}} (\rho_{\text{c}})$ ($L({\rho_{\text{c}}})$). By
comparing with the experimental data $\alpha_{\text{D}}=20.1\pm0.6$ fm$^3$, one
can extract a strong constraint of $L(\rho_{\text{c}})=48.6\pm7.9$ MeV.

The above constraint of $L(\rho_{\text{c}})=48.6\pm7.9$ MeV has been obtained
by neglecting the weak correlations between the $\alpha_{\text{D}}$ in $^{208}$Pb and
other $8$ macroscopic quantities. To test the robustness of this constraint
and to obtain a more precise constraint, for fixed $E_{\text{sym}}(\rho_{\text{c}})$ 
and $L(\rho_{\text{c}})$, we optimize all other $8$ parameters instead of simply 
fixing them at their default values in MSL0, by minimizing
the weighted sum of ${\chi}^2$ evaluated from the difference between SHF
prediction and the experimental data for some selected observables using the
simulated annealing technique~\cite{Agr05}. In particular, in the optimization, we chose
the following experimental data of spherical even-even nuclei, i.e., (i) the binding
energy $E_B$ of $^{16}$O,$^{40,48}$Ca, $^{56,68}$Ni, $^{88}$Sr, $^{90}$Zr,
$^{100,116,132}$Sn, $^{144}$Sm, $^{208}$Pb~\cite{Wan12}; (ii) the charge rms
radii $r_{\text{C}}$ of $^{16}$O, $^{40,48}$Ca, $^{56}$Ni, $^{88}$Sr, $^{90}$Zr,
$^{116,132}$Sn, $^{144}$Sm, $^{208}$Pb~\cite{Ang04,Blanc05}; (iii) the breathing
mode energy $E_{0}$ of $^{90}$Zr,$^{116}$Sn,$^{144}$Sm and $^{208}$Pb~\cite{You99}.
In the calculation of the breathing mode energy $E_{0}=\sqrt{m_1/m_{-1}}$, we
evaluate the inverse energy-weighted sum rule $m_{-1}$ with the constrained
Hartree-Fock (CHF) method and obtain the energy-weighted sum rule $m_{1}$ using
the double commutator sum rule~\cite{Boh79,Col04,Agr04,Sil06}. In addition, in
the optimization, we constrain the macroscopic parameters by requiring that (i)
the neutron $3p_{1/2}-3p_{3/2}$ energy level splitting in $^{208}$Pb should lie
in the range of $0.8-1.0$ MeV; (ii)$m_{s,0}^*$ should be greater than $m_{v,0}^*$
and here we set $m_{s,0}^*-m_{v,0}^* =0.1m$ ($m$ is nucleon mass in vacuum) to
be consistent with the extraction from global nucleon optical potentials
constrained by world data on nucleon-nucleus and (p,n) charge-exchange
reactions~\cite{XuC10}. As usual, in the optimization, we assign a theoretical
error $1.2$ MeV to E$_B$, $0.025$ fm to $r_C$ while use the experimental
error for breathing mode energy $E_0$ with a weight factor $0.08$, so that the
respective $\chi^2$ evaluated from each sort of experimental data is roughly
equal to the number of the corresponding data points~\cite{Bev03}.

Using the above optimization process, we evaluate the electric dipole polarizability
$\alpha_{\text{D}}$ in $^{208}$Pb as a function of $L(\rho_{\text{c}})$ for a
fixed $E_{\text{sym}}(\rho_{\text{c}})$, and the results are shown in Fig~\ref{AlphadLc} by
solid up-triangles (down-triangles) for $E_{\text{sym}}(\rho_{\text{c}})=26.45~(26.85)$ MeV.
It should be noted that for each pair of $E_{\text{sym}}(\rho_{\text{c}})$ and
$L(\rho_{\text{c}})$ with fixed values, the other $8$ macroscopic quantities have been
optimized accordingly as described above. It is interesting to see that the values of 
$\alpha_{\text{D}}$ with optimization are quite consistent with the results using the 
default values in MSL0 without optimization and only show a small upward shift compared 
with the latter. Comparing the results from optimization to the experimental data, one 
can obtain a constraint of $L(\rho_{\text{c}}) = 47.3 \pm 7.8$ MeV, which is again in 
good agreement with the constraint $L(\rho_{\text{c}})=48.6\pm7.9$ MeV extracted using
the default values in MSL0. These features demonstrate the validity of neglecting
the weak correlations between the $\alpha_{\text{D}}$ in $^{208}$Pb and other $8$
macroscopic quantities. The present constraint on $L(\rho_{\text{c}})$ further agrees
very well with the constraint $L(\rho_{\text{c}}) = 46.0 \pm 4.5$ MeV extracted from
analyzing the experimental data on the neutron skin thickness of Sn isotopes~\cite{Zha13}.
This is a very interesting finding since these two constraints are obtained from two
completely independent experimental observables.

In addition, using the constrained $E_{\text{sym}}(\rho_{\text{c}})$ and
$L(\rho_{\text{c}})$ together with the corresponding $8$ other optimized quantities, one can
easily extract the $E_{\text{sym}}(\rho)$ and $L(\rho )$ at saturation density
$\rho_0$, and the results are $E_{\text{sym}}(\rho_{\text{0}})=32.7\pm1.7$ MeV
and $L(\rho_{\text{0}})=47.1\pm17.7$ MeV, which are essentially consistent with
other constraints extracted from terrestrial experiments, astrophysical
observations, and theoretical calculations with controlled
uncertainties~\cite{Tsa12,Lat12,ChenLW12,LiBA12,Tew13}. Especially, our
present results agree surprisingly well with the constraint of
$E_{\text{sym}}({\rho _{0}}) = 31.2$-$34.3$ MeV and
$L({\rho _{0}}) = 36$-$55$ MeV (at 95\% confidence level) obtained
from analyzing the mass and radius of neutron stars~\cite{Ste12} as well as that of
$E_{\text{sym}}({\rho _{0}}) = 29.0$-$32.7$ MeV and
$L({\rho _{0}}) = 40.5$-$61.9$ MeV extracted from the
experimental, theoretical and observational analyses~\cite{Lat12}. Our
results are also in agreement with the constraint of
$E_{\text{sym}}({\rho _{0}}) = 32.0\pm1.8$ MeV and
$L({\rho _{0}}) =  43.1\pm15$ MeV from analyzing pygmy dipole
resonances (PDR) of $^{130,132}$Sn~\cite{Kli07} and that of
$E_{\text{sym}}({\rho _{0}}) = 32.3\pm1.3$ MeV and
$L({\rho _{0}}) =  64.8\pm15.7$ MeV from analyzing PDR of
$^{68}$Ni and $^{132}$Sn~\cite{Car10}. In addition, our results are further
consistent with the constraint of $E_{\text{sym}}({\rho _{0}}) = 32.3\pm1.0$
MeV and $L(\rho_{\text{0}}) = 45.2 \pm 10.0$ MeV extracted from analyzing
the experimental data of the binding energy difference of heavy isotope
pairs and the neutron skins of Sn isotopes~\cite{Zha13} as well as the
constraint of $E_{\text{sym}}({\rho _{0}}) = 32.5\pm0.5$
MeV and $L({\rho _{0}}) =  70\pm15$ MeV from a new finite-range
droplet model analysis of the nuclear mass~\cite{Mol12}.

\begin{figure*}[tbp]
\includegraphics[scale=0.5]{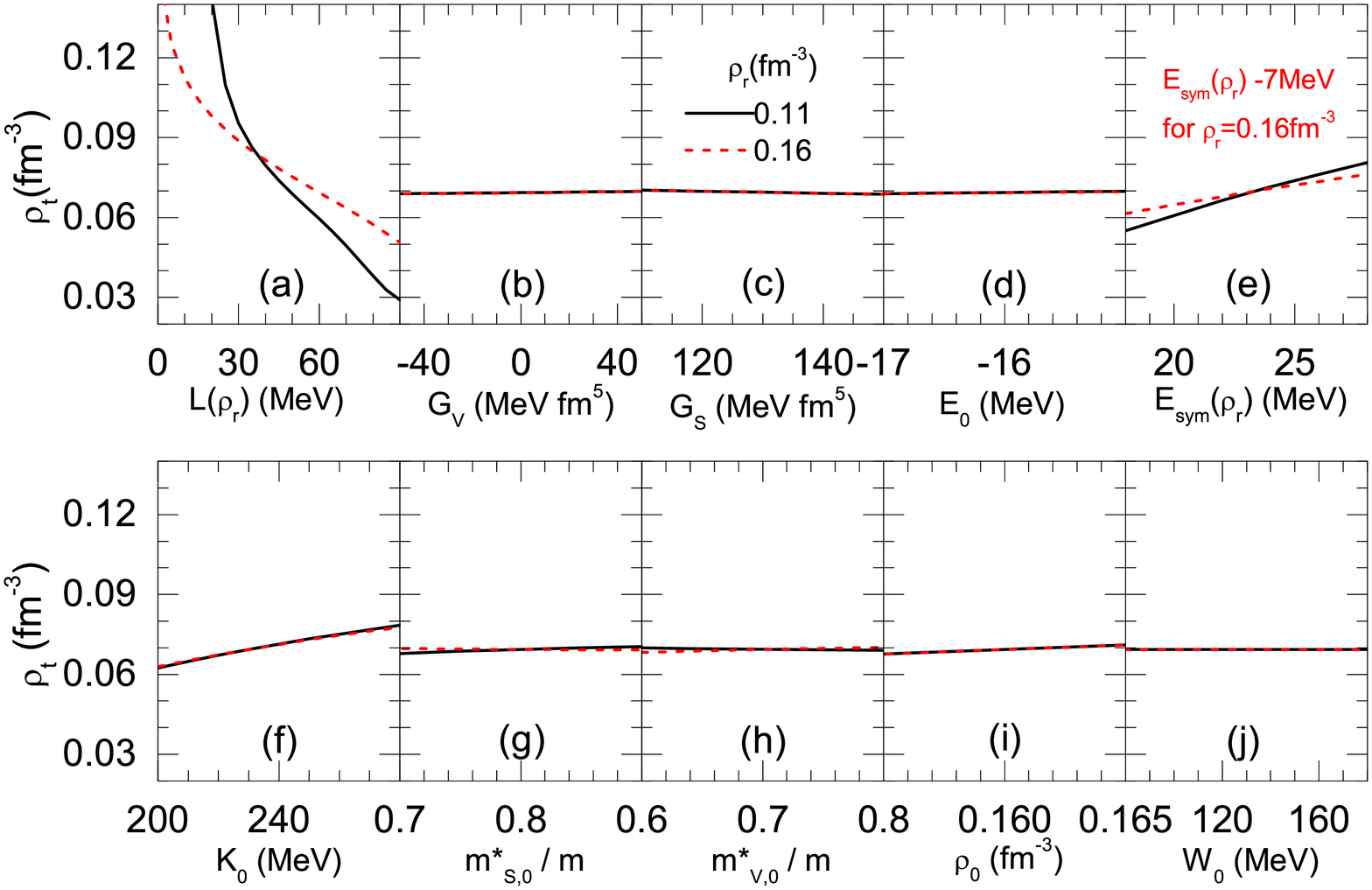}
\caption{(Color online) Same as Fig. \ref{AlphaD} but for the core-crust transition density $\rho_t$ in neutron stars.}
\label{Rhot}
\end{figure*}

Given that the neutron skin thickness ${\Delta}r_{np}$ of $^{208}$Pb is uniquely
fixed by the slope parameter $L(\rho_{\text{c}})$ at $\rho_{\text{c}}=0.11$
fm$^{-3}$~\cite{Zha13}, we can also extract a constraint ${\Delta}r_{np}=0.176\pm 0.027$
fm for $^{208}$Pb by using the optimized parameters together with
$E_{\text{sym}}(\rho_{\text{c}})=26.65\pm0.20$ MeV and
$L(\rho_{\text{c}}) = 47.3 \pm 7.8$ MeV. Our result is consistent with the
estimated range ${\Delta}r_{np}=0.165\pm(0.009)_{\text{expt}}\pm(0.013)_{\text{theor}}
\pm(0.021)_{\text{est}}$ fm in Ref.~\cite{Roc13} obtained by analyzing the
experimental $\alpha_{\text{D}}$ in $^{208}$Pb with an empirical range of
$E_{\text{sym}}(\rho_0)= 31\pm(2)_{\text{est}}$. One can see that
our present constraint on ${\Delta}r_{np}$ of $^{208}$Pb has higher precision, indicating
a more precise constraint on the symmetry energy at a subsaturation density is very
helpful to extract ${\Delta}r_{np}$ of $^{208}$Pb from the electric dipole
polarizability. Our result further agrees with the constraint $\Delta r_{np}=0.156^{+0.025}_{-0.021}$
fm obtained from the $^{208}$Pb dipole polarizability by using an empirical correlation
between $\alpha_{\text{D}}$ and ${\Delta}r_{np}$ of $^{208}$Pb ~\cite{Tam11}, the constraint
${\Delta}r_{np}=0.15\pm0.03({\text{stat.}})^{+0.01}_{-0.03}({\text{sys.}})$ fm extracted
very recently from coherent pion photoproduction cross sections~\cite{Tar14},
and within the experimental error bar the constraint $\Delta r_{np}=0.33^{+0.16}_{-0.18}$
fm extracted from the PREX at JLab~\cite{Abr12}.

Furthermore, it has been well established that the core-crust transition density 
$\rho_{\text{t}}$ in neutron stars, which plays a crucial role in neutron star 
properties~\cite{Lat04}, is strongly correlated with the density slope $L(\rho_0)$ 
of the symmetry energy (see, e.g., Ref.~\cite{XuJ09}). In particular, in Ref.~\cite{Che10}, 
the same correlation analysis method as in this work has been successfully applied 
to study the correlation between $\rho_{\text{t}}$ and the various macroscopic 
quantities, and indeed a strong correlation between $\rho_{\text{t}}$ and $L(\rho_0)$
has been found. As mentioned in Ref.~\cite{Zha13}, a similar strong correlation is 
also existed between $\rho_{\text{t}}$ and $L(\rho_{\text{c}})$, and this is demonstrated 
in Fig.~\ref{Rhot} which shows the same correlations as Fig. \ref{AlphaD} but for the 
core-crust transition density $\rho_{\text{t}}$ in neutron stars. Here, the transition 
density $\rho_{\text{t}}$ is calculated by using a dynamical approach
(see, e.g., Ref.~\cite{XuJ09}). One can see from Fig.~\ref{Rhot} that, for both 
$\rho_{\text{r}} =0.11$ and $0.16$ fm$^{-3}$, $\rho_{\text{t}}$ exhibits a strong 
correlation with $L({\rho_r})$, a weak dependence on $E_{\mathrm{sym}}(\rho_r)$ and 
$K_0$, but almost no sensitivity to other macroscopic parameters. Employing the 
optimized values for other macroscopic parameters as well as 
$E_{\text{sym}}(\rho_{\text{c}})=26.65\pm0.20$ MeV and 
$L(\rho_{\text{c}}) = 47.3 \pm 7.8$ MeV, we then obtain a value of 
$\rho_{\text{t}}= 0.084\pm0.009$ fm$^{-3}$, which agrees well with the empirical 
values~\cite{Lat04}.

\section{Summary and outlook}

In summary, we have demonstrated that the electric dipole polarizability
$\alpha_{\text{D}}$ in $^{208}$Pb is sensitive to both the magnitude
$E_{\text{sym}}(\rho_{\text{c}})$ and density slope $L(\rho_{\text{c}})$
of the symmetry energy at a subsaturation cross density
$\rho_{\text{c}} = 0.11$ fm$^{-3}$, and it decreases (increases) with
$E_{\text{sym}}(\rho_{\text{c}})$ ($L(\rho_{\text{c}})$), leading to
a positive correlation between $L(\rho_{\text{c}})$ and
$E_{\text{sym}}(\rho_{\text{c}})$ for a fixed value of $\alpha_{\text{D}}$
in $^{208}$Pb. Using the experimental value of $\alpha_{\text{D}}$
in $^{208}$Pb measured at RCNP and the very well-constrained range of
$E_{\text{sym}}(\rho_{\text{c}})$, we have obtained a strong constraint
on the slope parameter $L(\rho_{\text{c}}) = 47.3 \pm 7.8$ MeV. This constraint
is in surprisingly good agreement with the previous solely existing constraint
$L(\rho_{\text{c}}) = 46.0 \pm 4.5$ MeV from neutron skin data of Sn isotopes,
demonstrating the robustness of these constraints on the value of the
$L(\rho_{\text{c}})$ parameter.

The present constraint of $L(\rho_{\text{c}})$ further leads to
$E_{\text{sym}}(\rho_{\text{0}})=32.7\pm1.7$ MeV and
$L(\rho_{\text{0}}) = 47.1 \pm 17.7$ MeV for the symmetry energy at saturation
density, the neutron skin thickness $\Delta r_{np} = 0.176 \pm 0.027$ fm for
$^{208}$Pb, and $\rho_t = 0.084 \pm 0.009$ fm$^{-3}$ for the core-crust
transition density of neutron stars. These results are nicely consistent
with many other constraints extracted from terrestrial experiments, astrophysical
observations, and theoretical calculations with controlled uncertainties.

Our present results are based on the standard SHF energy density functional.
It will be interesting to see how the results change if different energy-density
functionals, e.g., the relativistic mean field model or the extended non-standard
SHF energy density functional, are applied. These works are in progress and will
be reported elsewhere.

\begin{acknowledgments}
We are grateful to Li-Gang Cao for helpful discussions on the
Skyrme-RPA code. This work was supported in part by the Major State Basic
Research Development Program (973 Program) in China under Contract Nos. 
2015CB856904 and 2013CB834405, the NNSF of China under Grant Nos. 11135011 
and 11275125, the ``Shu Guang" project supported by Shanghai Municipal 
Education Commission and Shanghai Education Development
Foundation, the Program for Professor of Special Appointment (Eastern Scholar)
at Shanghai Institutions of Higher Learning, and the Science and Technology
Commission of Shanghai Municipality (11DZ2260700).
\end{acknowledgments}


\begin{thebibliography}{99}

\bibitem{Lat04} J.M. Lattimer and M. Prakash, Science \textbf{304},
536 (2004); Phys. Rep. \textbf{442}, 109 (2007).

\bibitem{Ste05} A.W. Steiner, M. Prakash, J.M. Lattimer, and P.J. Ellis, Phys. Rep. \textbf{411}%
, 325 (2005).

\bibitem{Bar05} V. Baran, M. Colonna, V. Greco, and M. Di Toro, Phys. Rep. \textbf{410},
335 (2005).

\bibitem{LCK08} B.A. Li, L.W. Chen, and C.M. Ko, Phys. Rep. \textbf{464},
113 (2008).

\bibitem{Hor01b} C.J. Horowitz, S.J. Pollock, P.A. Souder, and R. Michaels, Phys. Rev. C \textbf{%
63}, 025501 (2001).

\bibitem{Sil05} T. Sil, M. Centelles, X. Vi$\tilde{\text{n}}$as, and J. Piekarewicz, Phys. Rev. C \textbf{71},
045502 (2005).

\bibitem{Wen09} D.H. Wen, B.A. Li, and L.W. Chen, Phys. Rev. Lett. \textbf{%
103}, 211102 (2009).

\bibitem{Zhe14} H. Zheng, Z. Zhang, and L.W. Chen, J. Cosmo. Astropart.
Phys. \textbf{08}, 011 (2014) [arXiv:1403.5134].

\bibitem{LiBAEPJA14} Topical Issue Nuclear Symmetry Energy edited by B.A. Li,
A. Ramos, G. Verde, I. Vidana, Eur. Phys. J. A \textbf{50}, (2014).

\bibitem{Tsa12} B.M. Tsang \textit{et al}., Phys. Rev. C \textbf{86}, 015803 (2012).

\bibitem{Lat12} J.M. Lattimer, Ann. Rev. Nucl. Part. Sci. \textbf{62}, 485 (2012).

\bibitem{ChenLW12} L.W. Chen, Nuclear Structure in China 2012:
Proceedings of the 14th National Conference on Nuclear
Structure in China (NSC2012) (World Scientific, Singapore,
2012), pp. 43-54 [arXiv:1212.0284].

\bibitem{LiBA12} B.A. Li \textit{et al}., J. Phys.: Conf. Series \textbf{413}, 012021 (2013) [arXiv:1212.1178].



\bibitem{Hor14} C.J. Horowitz \textit{et al}., J. Phys. G \textbf{41}, 093001 (2014).

\bibitem{Mye96} W.D. Myers and W.J. Swiatecki, Nucl. Phys. \textbf{A601}, 141 (1996).

\bibitem{Dan09} P. Danielewicz and J. Lee, Nucl. Phys. \textbf{A818}, 36 (2009).

\bibitem{Liu10} M. Liu, N. Wang, Z. X. Li, and F. S. Zhang, Phys. Rev. C \textbf{82}, 064306 (2010).

\bibitem{Mol12} P. M${\ddot{\text{o}}}$ller, W. D. Myers, H. Sagawa, and S. Yoshida, Phys. Rev. Lett. \textbf{108}, 052501 (2012).

\bibitem{LatX12} J.M. Lattimer and Y. Lim, Astrophys. J. \textbf{771}, 51 (2013)

\bibitem{War09} M. Warda, X. Vinas, X. Roca-Maza, and M. Centelles, Phys. Rev. C \textbf{80}, 024316 (2009).

\bibitem{Che10} L.W. Chen, C.M. Ko, B.A. Li, and J. Xu, Phys. Rev. C \textbf{%
82}, 024321 (2010).

\bibitem{Vin14} X. Vinas, M. Centelles, X. Roca-Maza, and M. Warda, Eur. Phys. J. A \textbf{50}, 27 (2014).

\bibitem{Sag07} H. Sagawa, S. Yoshida, X.-R. Zhou, K. Yako, and H. Sakai, Phys. Rev. C \textbf{76}, 024301 (2007).

\bibitem{Kli07} A. Klimkiewicz \textit{et al}., Phys. Rev. C \textbf{76}, 051603 (2007).

\bibitem{Car10} A. Carbone \textit{et al}., Phys. Rev. C \textbf{81}, 041301(R) (2010).

\bibitem{Rei10} P.--G. Reinhard and W. Nazarewicz, Phys. Rev. C \textbf{%
81}, 051303(R) (2010).

\bibitem{Pie12} J. Piekarewicz \textit{et al}., Phys. Rev. C \textbf{%
85}, 041302(R) (2012).

\bibitem{Pie14} J. Piekarewicz, Eur. Phys. J. A \textbf{50}, 25 (2014).

\bibitem{Col14} G. Col$\grave{\text{o}}$, U. Garg, and H. Sagawa, Eur. Phys. J. A \textbf{50}, 26 (2014).

\bibitem{Tam11}A. Tamii \textit{et al}., Phys. Rev. Lett. \textbf{107}, 062502 (2011).

\bibitem{Roc13} X. Roca-Maza \textit{et al}., Phys. Rev. C \textbf{88}, 024316 (2013).

\bibitem{Cen09} M. Centelles, X. Roca-Maza, X. Vin$\tilde{\text{a}}$s, and M. Warda, Phys. Rev. Lett. \textbf{102}, 122502 (2009)

\bibitem{Che11} L.W. Chen, Phys. Rev. C \textbf{83}, 044308 (2011).

\bibitem{Kha12} E. Khan, J. Margueron, and I. Vidana, Phys. Rev. Lett. \textbf{109}, 092501 (2012).

\bibitem{Fat14} F.J. Fattoyev, W.G. Newton, and B.A. Li, Phys. Rev. C \textbf{90}, 022801(R) (2014).

\bibitem{Zha13} Z. Zhang and L.W. Chen, Phys. Lett. \textbf{B726}, 234 (2013).

\bibitem{Dan14} P. Danielewicz and J. Lee, Nucl.\ Phys. \textbf{A922}, 1 (2014).

\bibitem{Tri08} L. Trippa, G. Col$\grave{\text{o}}$, and E. Vigezzi, Phys. Rev. C \textbf{77}, 061304(R) (2008).

\bibitem{Cao08} L.G. Cao and Z.Y. Ma, Chin. Phys. Lett. \textbf{25}, 1625 (2008).

\bibitem{Roc13a} X. Roca-Maza \textit{et al}., Phys. Rev. C \textbf{87}, 034301 (2013).

\bibitem{Bro13} B.A. Brown, Phys. Rev. Lett. \textbf{11}, 232502 (2013).


\bibitem{Wan13} N. Wang, L. Ou, and M. Liu, Phys. Rev. C \textbf{87}, 034327 (2013).

\bibitem{Cha97} E. Chabanat \textit{et al}., Nucl. Phys. \textbf{A627%
}, 710 (1997).

\bibitem{Fri86} J. Friedrich and P.-G. Reinhard, Phys. Rev. C
\textbf{33}, 335 (1986).

\bibitem{Klu09} P. Kl\"{u}pfel \textit{et al}., Phys. Rev. C
\textbf{79}, 034310 (2009).

\bibitem{Che12} L.W. Chen and J.Z. Gu, J. Phys. G \textbf{%
39}, 035104 (2012).

\bibitem{Che11a} L.W. Chen, Sci. China: Phys. Mech. Astro. \textbf{54} (Suppl. 1), s124 (2011).

\bibitem{Colo13} G. Col$\grave{\text{o}}$, L.G. Cao, N. Van Giai, and L. Capelli, Comput. Phys. Commun. \textbf{184}, 142 (2013).



\bibitem{Agr05} B.K. Agrawal, S. Shlomo, and V. Kim Au, Phys. Rev. C \textbf{72}, 014310 (2005).

\bibitem{Wan12} M. Wang \textit{et al}., Chin. Phys. C \textbf{36} (2012) 1287.


\bibitem{Ang04} I. Angeli, At. Data Nucl. Data. Tab. \textbf{87} (2004) 185.

\bibitem{Blanc05} F. Le Blanc \textit{et al}.,
 Phys. Rev. C \textbf{72}, (2005) 034305.

\bibitem{You99} D.H. Youngblood, H.L. Clark, and Y.W. Lui, Phys. Rev. Lett. \textbf{82}, 691 (1999).

\bibitem{Boh79} O. Bohigas, A.M. Lane, and J. Martorell, Phys. Rep. \textbf{51}, 267 (1979).

\bibitem{Col04} G. Col$\grave{\text{o}}$, N. Van Giai, J. Meyer, K. Bennaceur, and P. Bonche, Phys. Rev. C \textbf{70}, 024307 (2004).

\bibitem{Agr04} B.K. Agrawal and S. Shlomo, Phys. Rev. C \textbf{70}, 014308 (2004).

\bibitem{Sil06} T. Sil, S. Shlomo, B.K. Agrawal, and P.-G. Reinhard, Phys. Rev. C \textbf{73}, 034316 (2006).

\bibitem{XuC10} C. Xu, B.A. Li, and L.W. Chen, Phys. Rev. C \textbf{82}, 054607 (2010).

\bibitem{Bev03} P.R. Bevington and D.K. Robinson,
\textit{Data reduction and error analysis for physical sciences} (3ed.) (McGraw-Hill, New York, 2003), p. 71.


\bibitem{Tew13} I. Tews \textit{et al}., Phys. Rev. Lett. \textbf{%
110}, 032504 (2013).

\bibitem{Ste12} A.W. Steiner and S. Gandolfi, Phys. Rev. Lett. \textbf{108}, 081102 (2012).


\bibitem{Tar14} C.M. Tarbert \textit{et al}., Phys. Rev. Lett. \textbf{112}, 242502 (2014).

\bibitem{Abr12} S. Abrahamyan \textit{et al}., Phys. Rev. Lett
\textbf{108}, 112502 (2012).

\bibitem{XuJ09} J. Xu, L.W. Chen, B.A. Li, and H.R. Ma, Phys. Rev. C
\textbf{79}, 035802 (2009); Astrophys. J. \textbf{697}, 1549 (2009).

\end{thebibliography}
\end{document}